\documentclass[10pt,journal,comsoc,polish,english]{IEEEtran}

\usepackage[T1]{fontenc}
\usepackage[latin2,latin9,utf8]{inputenc}
\usepackage{array}
\usepackage{float}
\usepackage{mathtools}
\usepackage{amsmath}
\usepackage{amssymb}
\usepackage{graphicx}
\usepackage{color}

\makeatletter

\floatstyle{ruled}
\usepackage{bbm}

\pdfmapfile{+txfonts.map}

\newcommand{\Signal}{S}
\newcommand{\Noise}{\nu}
\newcommand{\Vis}{V}
\newcommand{\Vise}{V_e}
\newcommand{\Visd}{V_d}

\newcommand{\Ve}{V_e}
\newcommand{\Vd}{V_d}

\newcommand{\snrR}{\beta}
\newcommand{\SysOp}{{\cal N}}
\newcommand{\mj}{}

\makeatother

\usepackage{babel}
\begin{document}

\title{Quantum Fingerprinting over AWGN Channels with Power-Limited Optical Signals}
\author{Micha\l{}~Lipka, Marcin~Jarzyna, and Konrad~Banaszek, {\em Senior Member, IEEE}%
\thanks{Michał Lipka (e-mail: m.lipka@cent.uw.edu.pl), Marcin Jarzyna (e-mail: m.jarzyna@cent.uw.edu.pl) and Konrad Banaszek (e-mail: kbanasz@cent.uw.edu.pl) are with the Centre for Quantum Optical Technologies, Centre of New Technologies, University of Warsaw, Banacha 2c, 02-097 Warsaw, Poland. M.L.\ and K.B.\ are also with the Faculty of Physics, University of Warsaw, Pasteura 5, 02-093 Warsaw, Poland.}%
\thanks{This work is a part of the "Quantum Optical Technologies" project carried out within the  International Research Agendas programme of the Foundation for Polish Science co-financed by the European Union under the European Regional Development Fund. }}
%\markboth{IEEE JOURNAL ON SELECTED AREAS IN COMMUNICATIONS}%
%{Lipka \MakeLowercase{\textit{et al.}}: Quantum Fingerprinting over AWGN Channels with Power-Limited Optical Signals}%
\maketitle

\begin{abstract}
Quantum fingerprinting reduces communication complexity of determination whether two $n$-bit long inputs are equal or different in the simultaneous message passing model. Here we quantify the advantage of quantum fingerprinting over classical protocols when communication is carried out using optical signals with limited power and unrestricted bandwidth propagating over additive white Gaussian noise (AWGN) channels with power spectral density (PSD) much less than one photon per unit time and unit bandwidth. We identify a noise parameter whose order of magnitude separates near-noiseless quantum fingerprinting, with signal duration effectively independent of $n$, from a regime where the impact of AWGN is significant. In the latter case the signal duration is found to scale as $O(\sqrt{n})$, analogously to classical fingerprinting. However, the dependence of the signal duration on the AWGN PSD is starkly distinct, leading to quantum advantage in the form of a reduced multiplicative factor in $O(\sqrt{n})$ scaling.
\end{abstract}


\IEEEpeerreviewmaketitle

\section{Introduction}

Exploiting the quantum nature of physical signals used for information transmission enables new functionalities, such as quantum key distribution \cite{Gisin2002,Scarani2009,Xu2019}. It can also reduce communication complexity of certain distributed information-processing tasks. An example of the latter can be demonstrated in the simultaneous message passing model introduced by Yao \cite{Yao1979}. Suppose that two parties, Alice and Bob, receive inputs in the form of $n$-bit long strings $x,y \in \{0,1\}^n$. While they cannot communicate with each other, they are supposed to use as little communication as possible with a third party, the referee, to facilitate computation of a certain Boolean function $f(x,y)$. In the specific scenario of the equality problem, the function reads
\begin{equation}\label{Eq:Equality}
f(x,y)=\begin{cases}
1, & \mbox{if $x=y$},\\
0, & \mbox{if $x\neq y$},
\end{cases}
\end{equation}
which corresponds to a test whether the input strings are equal or different.
In order to reduce the amount of information transmitted to the referee, Alice and Bob can send only fingerprints of their inputs at the expense of tolerating a non-zero probability of error. Classically, the fingerprints have the form of bit strings shorter than inputs. If Alice and Bob do not have access to shared randomness, the fingerprints must be at least $O(\sqrt{n})$ bits long for an arbitrarily low probability of error
\cite{Ambainis1996,NewmanSzegedy,Babai1997}.
On the other hand, when quantum states are used to carry fingerprints, it is sufficient that Alice and Bob communicate to the referee $O(\log_2 n)$ qubits \cite{Buhrman2001,Buhrman2001ieee,Aaronson2004,Yao2003,Gavinsky2006}. Because according to Holevo's theorem \cite{Holevo1973,Holevo1998} a qubit can carry at most one bit of classical information, this presents a scaling advantage over classical fingerprinting. A key ingredient to attain this advantage is joint detection of quantum signals received from Alice and Bob by the referee.

Interestingly, quantum fingerprints can be efficiently generated as trains of coherent states of light
with joint detection implemented using optical interference and photon counting \cite{Arrazola2014,Lovitz2018}. Coherent states are routinely used in conventional optical communication, which facilitated recent experimental proof-of-principle demonstrations of quantum fingerprinting \cite{Xu2015,Guan2016}. This naturally leads to a question about the advantage of quantum fingerprinting over its classical counterpart in terms of physical resources required to transmit optical signals carrying fingerprints rather than by the number of bits or qubits that need to be communicated.

This paper presents an analysis of quantum fingerprinting  when optical signals sent from Alice and Bob to the referee are power-limited, but no restrictions on their bandwidth are in place. Our model includes contribution from background radiation described by additive white gaussian noise (AWGN). Motivated by recent studies of photon-starved communication \cite{BorosonSPIE2018,Zwolinski2018,BanaszekSPIE2019}, we consider regime when the noise power spectral density (PSD) $\nu$ expressed in photons per unit time per unit bandwidth is much less than one. The principal objective is to minimize the signal duration, which defines the transmission time required to execute the protocol.
We show that because the impact of AWGN becomes more severe with increasing signal bandwidth, there exists an optimal operating point that is determined by a combination of the input length $n$, the noise PSD $\nu$ and the desired probability of error $\varepsilon$ which is not to be exceeded when executing the protocol.

The obtained results are compared with a scenario when classical fingerprints are transmitted from Alice and Bob to the referee over optical channels with matching signal power and AWGN strength. This allows us to express quantum advantage in terms of reduction of the signal duration. We find that the performance of the quantum fingerprinting protocol changes qualitatively with increasing input size $n$. When $n\ll 2\Noise^{-1}\log[1/(2\varepsilon)]$, the effects of channel AWGN are insignificant and one remains close to the noiseless regime analyzed in \cite{Arrazola2014}. On the other hand, for sufficiently long inputs, when $n\gg 2\Noise^{-1}\log[1/(2\varepsilon)]$,
the transmission time for quantum fingerprints scales as $O(\sqrt{n})$, which is the same as in the classical scenario. However, the proportionality constant has a starkly distinct dependence on the noise PSD $\nu$. While in the classical scenario the noise PSD enters through a multiplicative factor $[\log_2(1+\nu^{-1})]^{-1}$, which follows directly from the Holevo capacity of an AWGN channel \cite{Giovannetti2014,Shapiro2009}, in the case of quantum fingerprinting the dependence is of the form $\sqrt{\nu}$. This difference becomes substantial for $\nu$ many orders below one photon per unit time and unit bandwidth, as is the case e.g.\ in space optical communication links \cite{HemmatiBook2006}.

This paper is organized as follows. Sec.~\ref{Sec:Optical} describes the optical layer of quantum fingerprinting based on coherent states of light. The complete quantum fingerprinting protocol is described in Sec.~\ref{Sec:Noiseless} for the noiseless case, and in Sec.~\ref{Sec:Testing} for a general AWGN scenario using the framework of  hypothesis testing.
Optimization of the operating point is discussed in Sec.~\ref{Sec:Optimization}. Sec.~\ref{Sec:Comparison} compares the performance of optimized quantum fingerprinting with classical protocols. Finally, Sec.~\ref{Sec:Conclusions} concludes the paper.

\section{Optical layer}
\label{Sec:Optical}

\begin{figure}
\centering

\includegraphics{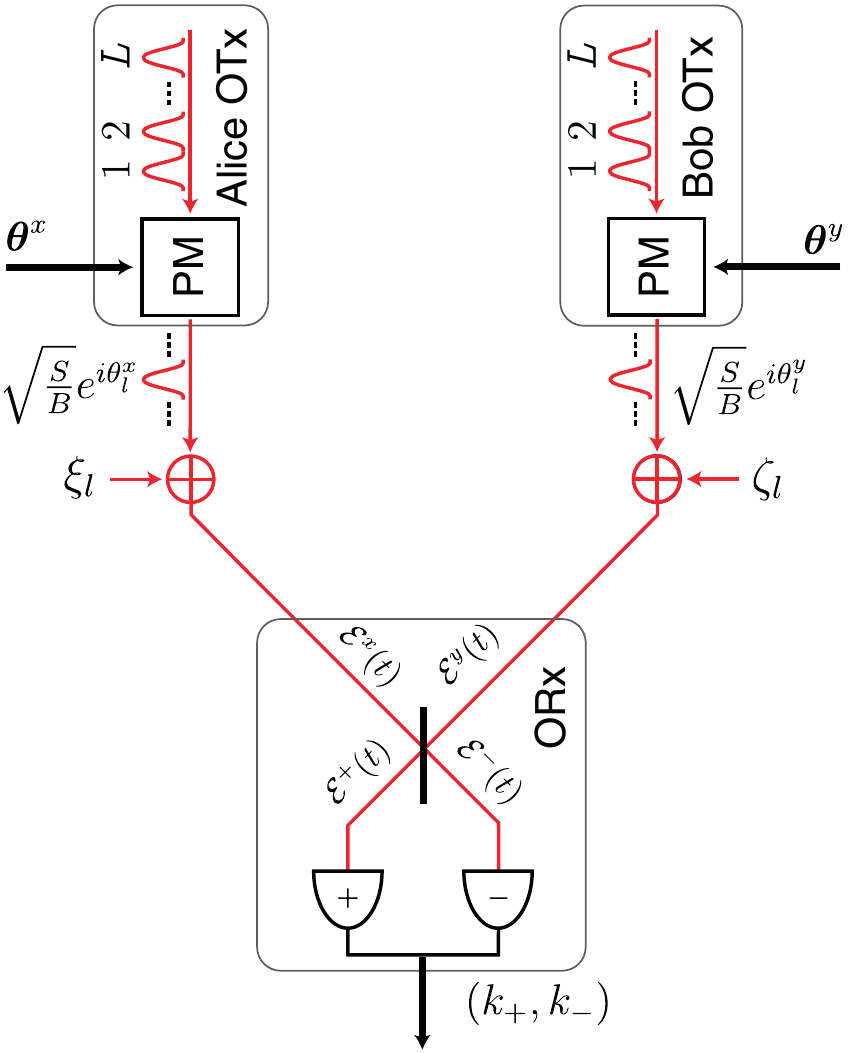}

\caption{Optical layer of the quantum fingerprinting protocol.
Alice and Bob use optical transmitters OTx which imprint phase $L$-tuples $\boldsymbol{\theta}^z = (\theta_1^z, \ldots, \theta_L^z)$ depending on inputs $z=x,y$  onto trains of $L$ light pulses using phase modulators PM. In the course of propagation, individual pulse amplitudes acquire random AWGN components $\xi_l$ and $\zeta_l$.
The optical receiver ORx used by the referee combines the received signals, described by time-dependent fields ${\cal E}^x(t)$ and ${\cal E}^y(t)$, on a balanced 50/50 beam splitter which produces superpositions
${\cal E}^{\pm}(t) = [{\cal E}^x(t)\pm{\cal E}^y(t)]/\sqrt{2}$. The output ports of the beam splitter are monitored by photon counting detectors which yield the total photocount numbers $k_+$ and $k_-$ registered over the signal duration.}

\label{Fig:Optical}

\end{figure}

Let us start with the description of the optical layer of the quantum fingerprinting protocol using coherent states proposed by Arrazola and L\"{u}tkenhaus \cite{Arrazola2014}. Alice and Bob use phase shift keying (PSK) to generate optical signals sent to the referee. As shown in Fig.~\ref{Fig:Optical}, each of the two signals is a train of $L$ optical pulses occupying consecutive temporal slots. A single pulse will be represented by a normalized mode function $u(s)$ parameterized with dimensionless time $s$.  It is assumed that the mode function is orthogonal to its replica displaced by any integer number $l$ of temporal slots:
\begin{equation}
\int_{-\infty}^{\infty}ds\,u^{\ast}(s-l)u(s)=\delta_{0l}, \quad l = \ldots, -1,0, 1, \ldots
\label{Eq:ModeFunction}
\end{equation}
For a modulation bandwidth $B$, the duration of a single slot is equal to $1/B$ and the physical time is $t=s/B$. Hence the overall duration of each of the signals is $L/B$. Note that in general the signal spectral support can exceed $B$ \cite{Essiambre2010}.

We will assume that the optical receiver used by the referee accepts only temporal modes matching those in the generated signals. Such selectivity can be achieved without any signal loss using the technique of quantum pulse gating \cite{BrechtEcksteinNJP2011,BrechtReddyPRX2015,Shahverdi2017,ReddyRaymerOptica2018}.
In this case, the optical fields ${\cal E}^{x}(t)$ and ${\cal E}^{y}(t)$ received by the referee respectively from Alice and Bob can be described by
\begin{equation}
\label{Eq:OpticalFields}
{\cal E}^{z}(t)  =\sqrt{B}\sum_{l=1}^{L}\alpha_{l}^{z}u(Bt-l), \quad z = x,y.
\end{equation}
Individual pulses are phase modulated by Alice and Bob according to $L$-tuples $\boldsymbol{\theta}^z = (\theta_1^z, \ldots, \theta_L^z)$, $z=x,y$, that depend on the input strings $x$ and $y$. The map $ z \mapsto \boldsymbol{\theta}^z$ will be specified in Sec.~\ref{Sec:Noiseless}. The complex amplitudes $\alpha_{l}^{x}$ and $\alpha_{l}^{y}$  in~(\ref{Eq:OpticalFields}) read
\begin{equation}
\alpha_{l}^{x}  =\sqrt{\frac{S}{B}}e^{i\theta_{l}^{x}}+\xi_{l},\quad
\alpha_{l}^{y}  = \sqrt{\frac{S}{B}} e^{i\theta_{l}^{y}}+\zeta_{l},
\label{Eq:alphajxy}
\end{equation}
where $\Signal$ is the optical power, in photons per unit time, of the signal received from either Alice or Bob.
Linear attenuation of the signal amplitude in the course of propagation can be taken into account in a straightforward manner by rescaling $\Signal$.
The complex variables $\xi_{l}$ and $\zeta_l$ describe contributions from AWGN acquired by the signals and will be assumed to have equal variance
\begin{equation}
\text{Var}[\xi_{l}]=\text{Var}[\zeta_{l}]=\Noise
\end{equation}
that specifies noise PSD expressed in photons per unit time per unit bandwidth. Because broadband noise is assumed, its contribution to field amplitudes $\alpha_{l}^{z}$ in~(\ref{Eq:alphajxy}) is independent of the modulation bandwidth $B$.

The referee brings the received optical signals to interfere on a balanced $50/50$ beam splitter. The fields
${\cal E}^{+}(t)$ and ${\cal E}^{-}(t)$ at the two $\pm$ output ports of the beam splitter, described by superpositions
\begin{equation}\label{Eq:OutputFields}
{\cal E}^{\pm}(t)=\frac{1}{\sqrt{2}}[{\cal E}^{x}(t)\pm{\cal E}^{y}(t)],
\end{equation}
are subsequently measured by a pair of photon counting detectors that return the total numbers of photocounts $k_+$ and $k_-$ registered over the entire signal duration. According to the semiclassical theory of photodetection \cite{MandelWolfSemiclPhot,Shapiro2009}, the probability distribution for the pair $(k_+, k_-)$ reads
\begin{equation}\label{eq:prob_comb}
p(k_{+},k_{-})=\mathbb{E}\left[e^{-I_+ } \frac{I_+^{k_+} }{k_+! }
e^{ - I_-}\frac{ I_-^{k_-}}{ k_-!}\right],
\end{equation}
where
\begin{equation}
I_{\pm}=\int_{-\infty}^\infty dt \,|{\cal E}^{\pm}(t)|^2
\label{Eq:Ipm}
\end{equation}
is the total optical energy incident on an individual detector over the signal duration and the expectation value $\mathbb{E}[\ldots]$ is calculated over all AWGN variables $\xi_{l}$ and $\zeta_l$, $l=1,\ldots,L$.
The characteristic function for the probability distribution $p(k_{+},k_{-})$ reads
\begin{align}
Z(\lambda_{+},\lambda_{-})&=\sum_{k_{+},k_{-}=0}^{\infty}e^{i\lambda_{+}k_{+}+i\lambda_{-}k_{-}}p(k_{+},k_{-})\nonumber \\
&=\mathbb{E}\left[\exp\left((e^{i\lambda_{+}}-1)I_{+}+(e^{i\lambda_{-}}-1)I_{-}\right)\right].
\label{Eq:Zlambda+lambda-}
%%=\mathbb{E}\left[\exp\left(\frac{1}{2}\sum_{j=1}^{m}
%%(e^{i\lambda_{+}}-1)|\alpha_{j}^{x}+\alpha_{j}^{y}|^{2}\right.\right.
%%\left.\left.+(e^{i\lambda_{-}}-1)|\alpha_{j}^{x}-\alpha_{j}^{y}|^{2}\vphantom{\sum_{j=1}^{m}}\right)\right],
\end{align}
The analysis will be carried out
%%%in the photon-starved regime, when $\Signal/B \ll 1$ and
for $\Noise \ll 1$. Further, terms of the order $O(\Noise L\Signal/B)$ and higher will be neglected. As shown in Appendix~\ref{Appendix:CharacteristicFunction}, under these assumptions the characteristic function after averaging over the noise variables can be recast as
\begin{align}
Z(\lambda_+, \lambda_-) &= \exp[(e^{i\lambda_+} -1) \mu (1+\Vis)] \nonumber \\ &\times \exp[(e^{i\lambda_-} -1) \mu (1-\Vis)],
\label{Eq:Zfinal}
\end{align}
where
\begin{equation}
\mu =  L (\Signal/B+\Noise)
\label{Eq:mudef}
\end{equation}
is the total number of photocounts generated on both the detectors by the noisy signal
coming from one sender, and
\begin{equation}
\Vis=\frac{1}{L(1+B\Noise/\Signal)}\sum_{l=1}^{L}\cos(\theta_{l}^{x}-\theta_{l}^{y}).
\label{Eq:vgen}
\end{equation}
has the physical interpretation of interference visibility. The characteristic function derived in~(\ref{Eq:Zfinal}) indicates Poissonian distributions for the photocount numbers $k_\pm$  with respective means $\mu (1\pm\Vis)$:
\begin{multline}
p(k_+, k_-|\Vis) = e^{-\mu (1+\Vis)}\frac{[\mu (1+\Vis)]^{k_+}}{k_+!} e^{-\mu (1-\Vis)}\frac{[\mu (1-\Vis)]^{k_-}}{k_-!}.
\label{Eq:pk+k-final}
\end{multline}
We have written explicitly the conditional dependence of the photocount statistics on the visibility $\Vis$, as this parameter contains information about the relation between the inputs $x$ and $y$.
The pair of photocount numbers $(k_+, k_-)$ produced by the detectors serves as the basis for testing by the referee whether the input strings $x$ and $y$ are different or equal.
%%The success rate of such a test depends on how much information

\section{Noiseless scenario}
\label{Sec:Noiseless}

\begin{figure}
\centering

\includegraphics{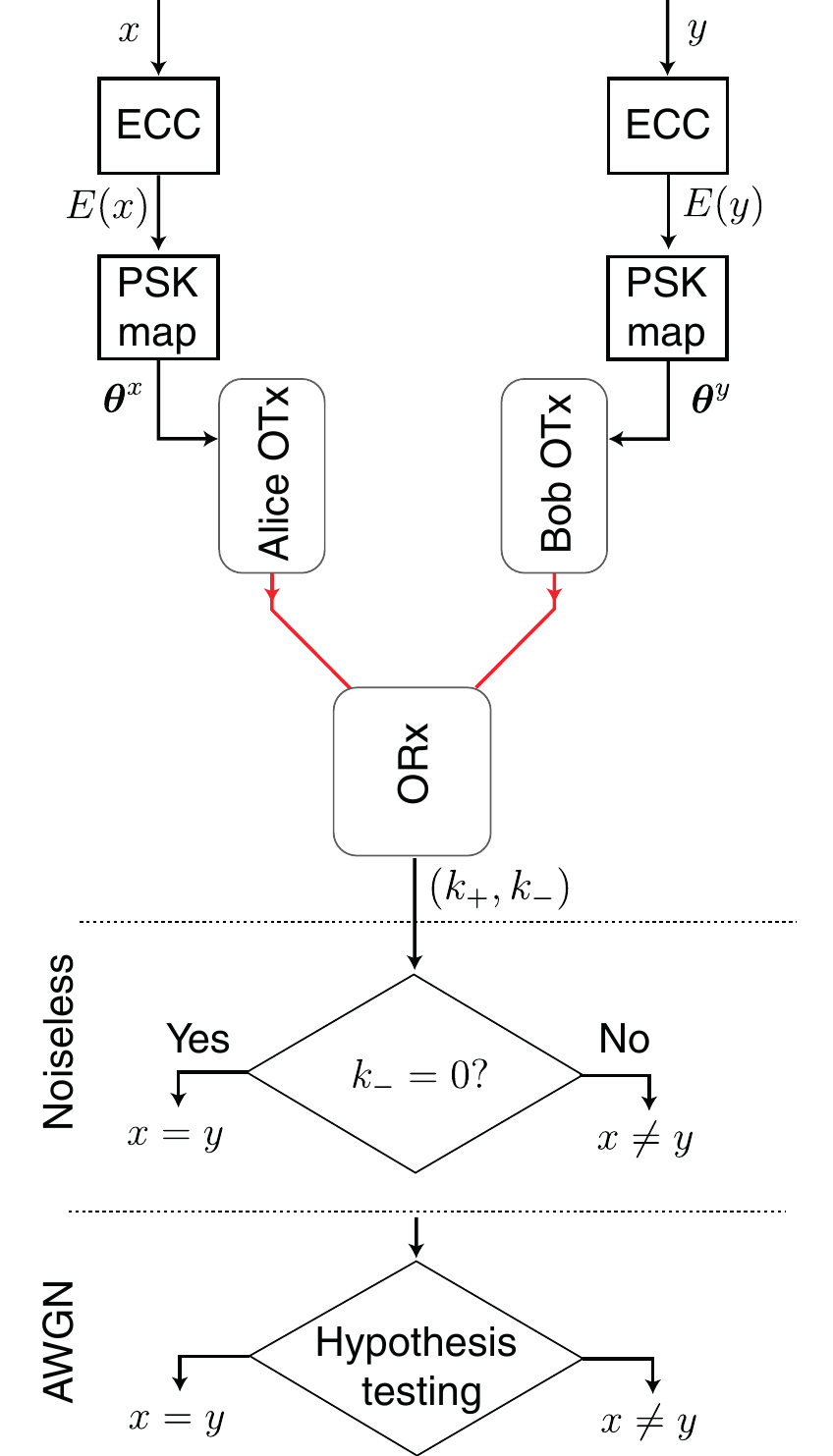}

\caption{Complete implementation of the quantum fingerprinting protocol based on coherent states of light. Inputs $x$ and $y$ are mapped onto codewords $E(x)$ and $E(y)$ using an error correcting code ECC. The codewords define via a PSK map phase $L$-tuples $\boldsymbol{\theta}^x$ and $\boldsymbol{\theta}^y$ that feed into optical transmitters OTx. The optical receiver ORx produces a pair of integers $k_+,k_-$ that serves as the basis for the equality test. In the noiseless case the test has the form of a check whether $k_-=0$ or not, whereas in the presence of noise a more complex test described in Sec.~\ref{Sec:Testing} is required.}

\label{Fig:Protocol}

\end{figure}

The optical layer described in the preceding section is used to implement the quantum fingerprinting protocol as shown in Fig.~\ref{Fig:Protocol}. The inputs $x$ and $y$ are mapped onto phase $L$-tuples $\boldsymbol{\theta}^x$ and $\boldsymbol{\theta}^y$ that define modulation of signals generated by Alice and Bob using optical transmitters OTx. Joint detection of these signals with an optical receiver ORx returns a pair of integers $(k_+,k_-)$ that is used by the referee to infer the value of the equality function defined in~(\ref{Eq:Equality}).

%\begin{figure}
%\centering
%\includegraphics{rysunki/psk}
%
%\caption{(a) Binary and (b) quaternary PSK maps.}
%
%\label{Fig:PSK}
%
%\end{figure}

We will begin with a discussion of a simplified scenario when there is no background noise, $\Noise = 0$.
In order to gain intuition about the workings of the fingerprinting protocol, suppose for a moment
that the binary input strings $x$ and $y$ of length $n$ are used directly to generate optical signals composed of $L=n$ pulses using a binary PSK map.
%shown in Fig.~\ref{Fig:PSK}(a).
In this setting, the two bit values $z_l=0,1$ are mapped onto phases $\theta^z_l = \pi z_l$, where $z$ stands for $x$ or $y$ and $l=1,\ldots, n$. For equal inputs, $x=y$, the two signals are identical,
completely destructive interference occurs at the `$-$' output port of the beam splitter, and ${\cal E}^{-}(t) = 0$ over the entire signal duration given absence of background noise.
%%%Consequently, the visibility (\ref{Eq:vgen}) is equal to one when zero background noise is assumed.
As a result, no photocounts can be registered by the detector monitoring the `$-$' port and $k_-=0$. Conversely, registering $k_- \ge 1$ photocounts heralds unambiguously that the inputs were different, $x\neq y$, as in this case ${\cal E}^{-}(t)$ is not identically equal to zero. However, because photon counting is a Poissonian process, it may happen that different strings will not produce any counts on the detector monitoring the `$-$' port. According to~(\ref{Eq:pk+k-final}) the probability of such an event is $p(k_- =0) = \exp[-\mu(1-V)]$. In the worst-case scenario, when the input strings differ at just one location, the visibility calculated according to~(\ref{Eq:vgen}) reads $\Vis =1-2/n$ and $p(k_- =0) =  \exp(-2\Signal/B)$. In order to keep this probability below a desired level, one would need to maintain sufficiently high ratio $\Signal/B$ which specifies the mean photon number per temporal slot. For power-limited signals this would imply an upper bound on the bandwidth $B$. Consequently, the entire signal duration given by $L/B = n/B$
would scale linearly with $n$.

Quantum fingerprinting offers dramatically improved performance compared to the simple scenario described above by using an error correcting code (ECC) to define the map $z \mapsto \boldsymbol{\theta}^z$, $z=x,y$ and exploiting bandwidth as a free resource. Specifically, consider a binary ECC $E:\{0,1\}^n \rightarrow \{0,1\}^m$, which guarantees that any two different inputs $x \neq y$ are mapped onto codewords $E(x)$ and $E(y)$ for which the Hamming distance satisfies
\begin{equation}
D \bigl( E(x), E(y) \bigr) = \sum_{j=1}^m E_j(x) \oplus E_j(y)
\ge m\delta.
\label{Eq:MinHamming}
\end{equation}
Here $\delta \in [0,1/2[$ is a constant specifying the minimum relative Hamming distance between any two different codewords. It will be assumed that the ECC $E$ operates at the asymptotic Gilbert-Varshamov bound given by \cite{Lint1987}
\begin{equation}
\frac{n}{m} = r(\delta) = 1 - H_2(\delta),
\end{equation}
%and plotted in Fig.~\ref{Fig:GVbound}.
where $H_2(x) = -x \log_2 x - (1-x) \log_2 (1-x)$ is the binary entropy. \mj{
There exist efficient ECCs operating close to the Gilbert-Varshamov bound, such as the random Toeplitz matrix ECC employed in a recent experimental demonstration of quantum fingerprinting \cite{Guan2016}.}

%\begin{figure}
%\centering
%\includegraphics{wykresy/rgv}
%
%\caption{Asymptotic Gilbert-Varshamov bound for the ECC rate $r(\delta)$ as a function of the minimum relative Hamming distance $\delta$ between any two distinct codewords.}
%
%
%\label{Fig:GVbound}
%
%\end{figure}

The codewords $E(x)$ and $E(y)$ are mapped onto $L$-tuples of phases $\boldsymbol{\theta}^x$ and $\boldsymbol{\theta}^y$ that are used to modulate optical signals. We shall take $L=m/2$ and employ a quadrature PSK map so that an individual phase depends on a block of two consecutive codeword bits according to
\begin{equation}
\theta^z_l = \pi E_{2l-1}(z)  + {\frac{\pi}{2}} [E_{2l-1}(z) \oplus E_{2l}(z)], \quad z=x,y,
\end{equation}
%as illustrated in Fig.~\ref{Fig:PSK}(b).
where $l=1,\ldots, L=m/2$.
Compared to binary PSK, quadrature PSK allows for a two-fold reduction of the pulse train length without altering otherwise the performance of the protocol \cite{Lovitz2018}. This would no longer be the case for higher PSK constellations. Calculation of the interference visibility (\ref{Eq:vgen}) is aided by the following straightforward observation:
\begin{multline}
\cos(\theta^x_l - \theta^y_l)=1 - E_{2l-1}(x) \oplus E_{2l-1}(y) - E_{2l}(x) \oplus E_{2l}(y).
\end{multline}
Assuming absence of noise, one obtains:
\begin{align}
\Vis &= \frac{1}{L} \sum_{l=1}^{L} \cos(\theta^x_l - \theta^y_l) = 1- \frac{2}{m} \sum_{j=1}^{m} E_j(x) \oplus E_j(y) = \nonumber \\
 &= 1 - \frac{2}{m} D \bigl( E(x), E(y) \bigr) \le 1-2\delta,
 \end{align}
where in the last step~(\ref{Eq:MinHamming}) has been used.
The probability of obtaining $k_-=0$ for different inputs, $x \neq y$, is consequently upper bounded by
$\exp(-2 \delta L S/ B)$. % = \exp\{-2\delta n S/ [B r(\delta)]\}$.
Given that $L/B$ specifies the signal duration, it is now possible to execute the quantum optical fingerprinting protocol in a constant time by increasing the modulation bandwidth in line with $L$ which grows with the input size $n$ as $L=n/[2r(\delta)]$. Without any bandwidth limitations, it is optimal to approach  $\delta \rightarrow 1/2$. In this limit the code rate $r(\delta)\rightarrow 0$ and the number of temporal slots $L \rightarrow \infty$. With unlimited bandwidth these slots can be accommodated in a constant time $L/B$.

It is worth noting that the ECC is used in quantum fingerprinting  {\em not} to ensure faithful recovery of the messages fed into the communication channel, but rather to augment differences between received optical signals in order to guarantee sufficiently low interference visibility when $x\neq y$ which results in photocounts on the `$-$' detector.

\section{Hypothesis testing}
\label{Sec:Testing}

In the remainder of the paper, the fingerprinting protocol will be required to operate at or below a desired average probability of error $\varepsilon$ for the equality test, assuming equiprobable hypotheses of equal and different inputs, and considering for the latter hypothesis the worst-case scenario of the minimum relative Hamming distance $\delta$ between the codewords. The objective will be to minimize the overall duration of signals sent by Alice and Bob given by $L/B$. For a fixed signal power $S$, the signal duration can be equivalently characterized by the signal optical energy expressed as the mean photon number received from Alice or Bob that is equal to $N_Q = SL/B$. In the noiseless case discussed in the preceding section, assuming unlimited bandwidth and taking $\delta\rightarrow 1/2$ yields the average probability of error equal to $\varepsilon = \exp(- N_Q)/2$, which can be recast as:
\begin{equation}
\label{Eq:NQNoiseless}
N_Q = \log[1/(2\varepsilon)], \qquad \Noise = 0.
\end{equation}
This expression is independent of the input length $n$ implying constant signal duration.
As expected,  a lower probability of error requires higher photon number or, equivalently for power-limited signals, longer transmission time.

The above analysis becomes much more nuanced when background noise is present. First, the simple test based on whether $k_-=0$ or not no longer guarantees minimum probability of error. Second, while in the noiseless case there was no penalty for increasing the bandwidth in order to accommodate more temporal slots within a constant transmission time, higher bandwidth boosts the AWGN contribution to the received signals, which may make the equality test based on interference visibility increasingly more difficult.

\begin{figure}
\centering
\includegraphics{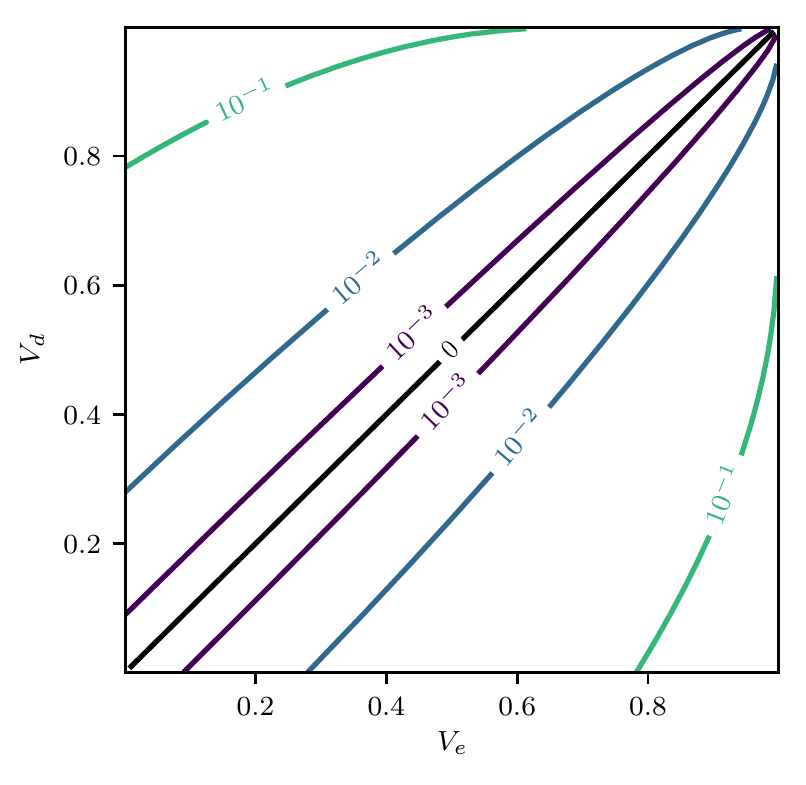}

\caption{Chernoff information per count $\textsf{C}(\Vise,\Visd)$ as a function of interference visibilities $\Vise$ and $\Visd$ corresponding respectively to hypotheses of equal and different inputs.}

\label{Fig:ChernoffPerClick}

\end{figure}

In the general scenario with background noise, the visibilities corresponding to hypotheses of equal and different inputs, assuming for the latter the worst-case scenario with the minimum relative Hamming distance $\delta$, are given respectively by
\begin{equation}
\Vise = \frac{1}{1+ B\Noise/\Signal}, \qquad \Visd = \frac{1-2\delta}{1+ B\Noise/\Signal}.
\label{Eq:VeVd}
\end{equation}
The referee needs to decide whether the pair of integers $(k_+, k_-)$ produced by the joint detection of optical signals received from Alice and Bob was generated by the probability distribution $p_e(k_+, k_-|\Vise)$ or $p_d(k_+, k_-|\Visd)$. We will use the Neyman-Pearson criterion for \emph{a priori} equiprobable hypotheses, which yields the decision rule
\begin{align}
p(k_+,k_-|\Vise) & > p(k_+,k_-|\Visd) : & x & = y \nonumber \\
p(k_+,k_-|\Vise) & < p(k_+,k_-|\Visd): & x & \neq y \nonumber
\end{align}
and a random draw when $p(k_+,k_-|\Vise) = p(k_+,k_-|\Visd)$. The probability of error for such a test is upper bounded by the Chernoff bound \cite{Cover1991}
\begin{equation}
\varepsilon \le \frac{1}{2} \exp[-C(V_e, V_d ; \mu)],
\label{Eq:ChernoffBound}
\end{equation}
where $C(V_e, V_d ; \mu)$ is Chernoff information given by
\begin{align}
C(V_e, V_d ; \mu) &= - \min_{0 \le \lambda \le 1} \log \Bigg\lbrace \sum_{k_+,k_- =0}^{\infty} [p(k_+, k_-|\Vise)]^\lambda \nonumber
\\ &\times [p(k_+, k_-|\Visd)]^{1-\lambda} \Bigg\rbrace.
\end{align}
As specified in~(\ref{Eq:pk+k-final}), the joint probability distributions $p(k_+, k_-|\Vise)$ and $p(k_+, k_-|\Visd)$ are products of Poissonian distributions with respective means $\mu(1\pm \Vise)$ and $\mu(1\pm \Visd)$. In such a case, Chernoff information is proportional to the total photocount number $2\mu$,
\begin{equation}
C(V_e, V_d ; \mu) = 2 \mu \textsf{C}(\Vise,\Visd).
\label{Eq:Chernoffproportional}
\end{equation}
The multiplicative factor $\textsf{C}(\Vise,\Visd)$ can be interpreted as {\em Chernoff information per count} and is given by the expression
\begin{align}
\textsf{C}(\Vise,\Visd)&=1-\frac{1}{2}\min_{0\le\lambda\le1}[(1+\Vise)^{\lambda}(1+\Visd)^{1-\lambda} \nonumber \\
&+(1-\Vise)^{\lambda}(1-\Visd)^{1-\lambda}].
\label{Eq:ChernoffPerClick}
\end{align}
Fig.~\ref{Fig:ChernoffPerClick} depicts $\textsf{C}(\Vise,\Visd)$ as a function of visibilities $\Vise$ and $\Visd$ for $0 \le \Vise, \Visd \le 1$. In this range,
Chernoff information per count attains maximum at $\textsf{C}(1,0)= \textsf{C}(0,1) =1/2$ and becomes zero for equal arguments. It will be useful to note that for a fixed $\Vise$, $\textsf{C}(\Vise,\Visd)$ is a decreasing function on an interval $\Visd \in [0, \Vise]$. The intuition behind this is that the closer $\Visd$ becomes to $\Vise$, the more difficult it is to discriminate between the two visibilities based on the photocount statistics. As derived in Appendix~\ref{Sec:AppB},
for $\Vise,\Visd \ll 1$ the Chernoff information per count is well approximated by the expression
\begin{equation}
\textsf{C}(\Vise,\Visd) \approx \frac{1}{8} (\Vise-\Visd)^2.
\label{Eq:ChernoffLowVisibility}
\end{equation}
This simple formula will greatly simplify the analysis of the performance of the quantum fingerprinting protocol in the limit of large input size $n$.

\section{Optimization}
\label{Sec:Optimization}

The task now is to identify the operating point achieving the minimum transmission time equal to $L/B$ or
equivalently---owing to the power constraint---the number of signal photons $N_Q = \Signal L/B$ that need to be received by the referee from Alice and Bob. The operating point depends on the input bit string length $n$, the noise strength $\nu$ and the desired average probability of error $\varepsilon$ which is not to be exceeded. It will be convenient to use as independent variables in the optimization problem the minimum relative Hamming distance $\delta$ of the ECC used in the protocol and the rescaled bandwidth
\begin{equation}
\snrR = \frac{B\Noise}{\Signal}.
\label{Eq:snrR}
\end{equation}
Note that the inverse $\snrR^{-1}$ specifies the signal-to-noise ratio. The range of the variables is $0 \le \delta < 1/2$ and $\beta > 0$.

Transforming the Chernoff bound (\ref{Eq:ChernoffBound}) with the help of definitions (\ref{Eq:mudef}), (\ref{Eq:VeVd}), and (\ref{Eq:Chernoffproportional}) implies that the photon number
\begin{equation}
N_Q \ge \frac{\log[1/(2\varepsilon)]}{2 (1+ \snrR) \textsf{C} \left( \frac{1}{1+\snrR}, \frac{1-2\delta}{1+\snrR}\right)}
\label{Eq:NQChernoff}
\end{equation}
is sufficient to ensure operation below a desired error probability $\varepsilon$. At the same time, the transmission time must be sufficiently long to accommodate $L=n/[2r(\delta)]$ temporal slots each of duration $1/B = \Noise/(\snrR\Signal)$. This condition translated for the number of received signal photons yields the inequality
\begin{equation}
N_Q \ge \frac{\Signal L}{B} = \frac{n\Noise/2}{\snrR r(\delta)}.
\label{Eq:NRCoderade}
\end{equation}
For a fixed $\beta$ the expressions on the right hand sides of~(\ref{Eq:NQChernoff}) and (\ref{Eq:NRCoderade}) exhibit opposite monotonicity as functions of $\delta$ over the interval $0 \le \delta < 1/2$. This is because in~(\ref{Eq:NQChernoff}), Chernoff information per count $\textsf{C} \left( \frac{1}{1+\snrR}, \frac{1-2\delta}{1+\snrR}\right)$ is monotonically increasing in $\delta$ as noted in Sec.~\ref{Sec:Testing}, while the code rate $r(\delta)$ in the denominator of~(\ref{Eq:NRCoderade}) is monotonically decreasing in $\delta$. % as illustrated in Fig.~\ref{Fig:GVbound}.
Consequently, if one seeks minimum $N_Q$ that satisfies both inequalities (\ref{Eq:NQChernoff}) and (\ref{Eq:NRCoderade}), it is sufficient to consider the case when the expressions on the right hand sides of these inequalities are equal to each other. This yields an implicit relation between $\snrR$ and $\delta$ in the form
\begin{equation}
\frac{\snrR r(\delta)}{2 (1+ \snrR) \textsf{C} \left( \frac{1}{1+\snrR}, \frac{1-2\delta}{1+\snrR}\right)}
= \SysOp,
\label{Eq:snrR-delta}
\end{equation}
where
\begin{equation}
\label{Eq:NoiseParameter}
\SysOp = \frac{n\Noise/2}{\log[1/(2\varepsilon)]}.
\end{equation}
The ratio defined in~(\ref{Eq:NoiseParameter}) admits a simple interpretation. The enumerator is the total number of noise photons if the inputs were mapped onto quadrature PSK signals without an ECC. The denominator is the number of signal photons required to implement the quantum fingerprinting protocol for the desired probability of error $\varepsilon$ in the noiseless scenario. Hence $\SysOp$ can serve as a simple estimate of how severely the background noise would impact the protocol designed for the noiseless case. In the following we will  refer to $\SysOp$ as the {\em noise parameter}.

\begin{figure}
\centering
\includegraphics{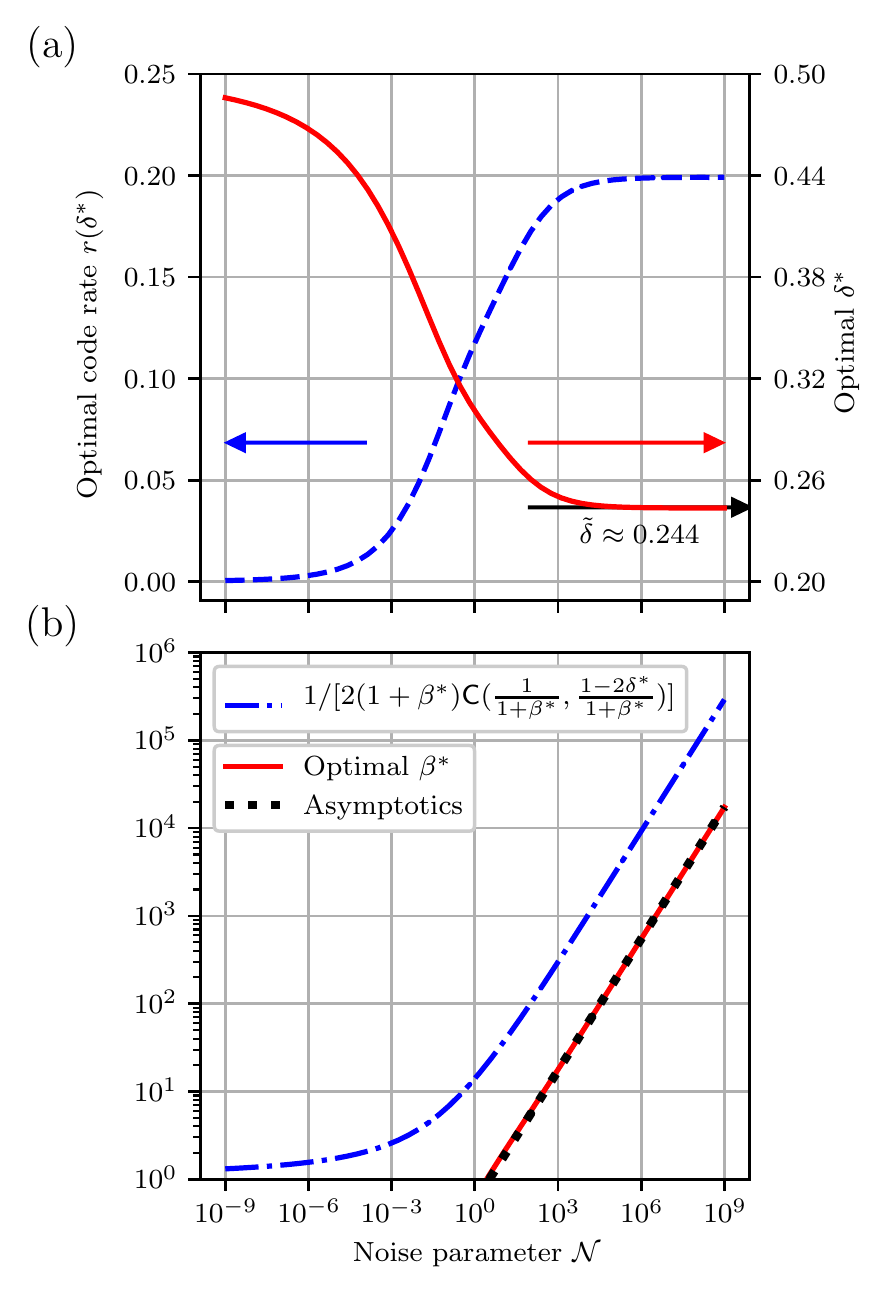}

\caption{(a) Optimal Hamming distance $\delta^\ast$ (solid line, right scale) and the corresponding code rate $r(\delta^\ast)$ (dashed line, left scale) minimizing the signal duration, or equivalently the signal photon number, as a function of the noise parameter ${\SysOp}$ defined in~(\ref{Eq:NoiseParameter}).
(b) Optimal rescaled bandwidth $\beta^\ast$ (solid line) compared with the asymptotic expression (dotted line) derived in~(\ref{Eq:snrRasympt}). The dash-dotted line depicts the proportionality factor between the minimum signal photon number and $\log_2[1/(2\varepsilon)]$, where $\varepsilon$ is the desired average probability of error.}

\label{Fig:Optimal}

\end{figure}

Equation~(\ref{Eq:snrR-delta}) provides a relation between $\snrR$ and $\delta$ that can be used to reduce the number of independent optimization variables to one and to find the optimum operating point by minimizing the right hand side of either~(\ref{Eq:NQChernoff}) or~(\ref{Eq:NRCoderade}) over the remaining variable. Fig.~\ref{Fig:Optimal} depicts numerically found optimal $\delta^\ast$ and the corresponding $\snrR^\ast$ as a function of the noise parameter $\SysOp$.
Two operating regimes can be identified depending on the order of magnitude of $\SysOp$.
When $\SysOp \ll 1$ it is possible to attain $\delta^\ast \approx 1/2$ and $\snrR^\ast \ll 1$. This corresponds to large ECC expansion with the code rate approaching $r(\delta^\ast) \approx 0$, as shown in Fig.~\ref{Fig:Optimal}(a). In this regime the minimum photon number $N_Q^\ast$ can be conveniently calculated using the right hand side of~(\ref{Eq:NQChernoff}) as a product of $\log[1/(2\varepsilon)]$ and a factor $1/\left[ 2 (1+\beta^\ast) \textsf{C} \left( \frac{1}{1+\snrR^\ast}, \frac{1-2\delta^\ast}{1+\snrR^\ast}\right)\right]$, plotted in Fig.~\ref{Fig:Optimal}(b). For ${\SysOp}\le 10^{-1}$ this factor remains between $1$ and $6.6$. Thus the fingerprinting protocol requires transmission time that depends primarily on the desired probability of error and the minimum number of signal photons
\begin{equation}
N_Q^\ast \approx \log[1/(2\varepsilon)] , \qquad \SysOp \ll 1.
\end{equation}
is within one order of magnitude the same as in the noiseless scenario.

Fig.~\ref{Fig:Optimal}(b) indicates that in the opposite regime, when $\SysOp \gg 1$, the rescaled bandwidth becomes $\beta \gg 1$, which corresponds to low signal-to-noise ratio. This allows one to apply the low-visibility approximation of the Chernoff information per count according to~(\ref{Eq:ChernoffLowVisibility}). This approximation expressed in presently used variables takes the form:
\begin{equation}
\label{Eq:Chernoffapprox1/(1+snrR)}
\textsf{C}\left( \frac{1}{1+\snrR}, \frac{1-2\delta}{1+\snrR}\right) \approx \frac{\delta^2}{2(1+\beta)^2}.
\end{equation}
Using the above closed formula in~(\ref{Eq:snrR-delta}) and solving it with respect to $\snrR$ yields $\snrR = \sqrt{\SysOp \delta^2 /r(\delta)+1/4} - 1/2 \approx \sqrt{\SysOp \delta^2 /r(\delta)}$, where the second approximate expression can be applied when $\snrR \gg 1$. Inserting the latter expression for $\snrR$ into the right hand side of~(\ref{Eq:NRCoderade}) yields $n\Noise/[2\sqrt{\SysOp \delta^2 r(\delta)}]$ that needs to be optimized over $\delta$. The product $\delta^2 r(\delta)$ appearing in the denominator has a single maximum over the interval $0 \le \delta < 1/2$ at the argument whose numerically found value  is equal to $\tilde{\delta} \approx 0.244$. As seen in Fig.~\ref{Fig:Optimal}(a), this value agrees very well with the results of numerical optimization for $\SysOp \gg 1$. Consequently, one can take
\begin{equation}
\snrR^\ast \approx \sqrt{\SysOp \tilde{\delta}^2 /r(\tilde{\delta})},\qquad \SysOp \gg 1,
\label{Eq:snrRasympt}
\end{equation}
and express the minimum photon number  using the right hand side of~(\ref{Eq:NRCoderade}) as:
\begin{equation}
N_Q^\ast \approx 6.51 \sqrt{n \Noise {\textstyle \log[{1}/{(2\varepsilon)}]} },
\qquad \SysOp \gg 1,
\label{Eq:NQOptLargeNoise}
\end{equation}
where the numerical multiplicative factor is given by the inverse of ${\sqrt{2\tilde{\delta}^2 r(\tilde{\delta})}} \approx 0.154$.

\section{Comparison}
\label{Sec:Comparison}

The performance of the optimized quantum fingerprinting protocol can be compared directly with a scenario when optical channels are used to transmit classical fingerprints of inputs $x$ and $y$. Based on results obtained by Babai and Kimmel \cite{Babai1997} one can specify a classical protocol that uses fingerprints of length
\begin{equation}
\label{Eq:ICdef}
I_C = 2 \sqrt{n} \left\lceil \frac{1}{2} \log_2 \frac{1}{\varepsilon} \right\rceil
\end{equation}
bits each. It is also possible to devise a lower bound on the classical fingerprint length in the form
\cite{Guan2016}
\begin{equation}
\label{Eq:IBdef}
I_B = \sqrt{\frac{n}{2\log 2}} \left( \frac{1}{2} - \sqrt{\varepsilon} \right) -\frac{1}{2}.
\end{equation}
It is worth noting that $I_B$ retains $O(\sqrt{n})$ scaling in the limit $\varepsilon \rightarrow 0$, which suggests that this bound is not tight. When the desired probability of error is equal to zero, it should be necessary to transmit entire inputs, leading to a breakdown of $O(\sqrt{n})$ scaling. This is the case of $I_C$ defined in~(\ref{Eq:ICdef}).

The maximum attainable rate $R$ in bits per unit time for transmission of classical information over an AWGN channel, allowing for the most general detection strategies, follows from the Holevo capacity and is given by \cite{Giovannetti2014}
\begin{equation}
\label{Eq:InfRate}
R = B[g(\Signal/B + \Noise) - g(\Noise)],
\end{equation}
where
\begin{equation}
g(x) = (x+1) \log_2 (x+1) - x \log_2 x
\end{equation}
is the entropy of a thermal state of a quantized harmonic oscillator with the mean number of excitations equal to $x$. For a given signal power $\Signal$ and noise PSD $\Noise$ the information rate is maximized in the limit $B \rightarrow \infty$. The first term in~(\ref{Eq:InfRate}) can be then expanded around $\Noise$ up to the first order in $\Signal/B$. This yields $R = \Signal g'(\Noise)$, where $g'(x) = \log_2 (1+x^{-1})$ is the first derivative of $g(x)$. The coefficient $g'(\Noise)$ has the interpretation of photon information efficiency (PIE), which specifies how many bits of information can be encoded in one photon \cite{Guha2011,BanaszekSPIE2019}. Consequently, $I_C$ and $I_B$ defined respectively in~(\ref{Eq:ICdef}) and (\ref{Eq:IBdef}) divided by PIE characterize the performance of classical fingerprinting in terms of total photon numbers carried by optical signals sent from Alice and Bob to the referee. Specifically,
\begin{equation}
N_C = \frac{I_C}{\log_2(1+ \Noise^{-1})} = \frac{2 \sqrt{n} }{\log_2(1+ \Noise^{-1})}\left\lceil \frac{1}{2} \log_2 \frac{1}{\varepsilon} \right\rceil
\label{Eq:NC}
\end{equation}
is sufficient to implement a constructive classical fingerprinting protocol, and
\begin{align}
N_B &= \frac{I_B}{\log_2(1+ \Noise^{-1})} \\
&= \frac{1}{\log_2(1+ \Noise^{-1})} \left[\sqrt{\frac{n}{2\log 2}} \left( \frac{1}{2} - \sqrt{\varepsilon} \right) -\frac{1}{2}\right] \nonumber
\label{Eq:NB}
\end{align}
defines a lower bound on the total signal photon number required by any classical fingerprinting protocol.

\begin{figure}
\centering
\includegraphics{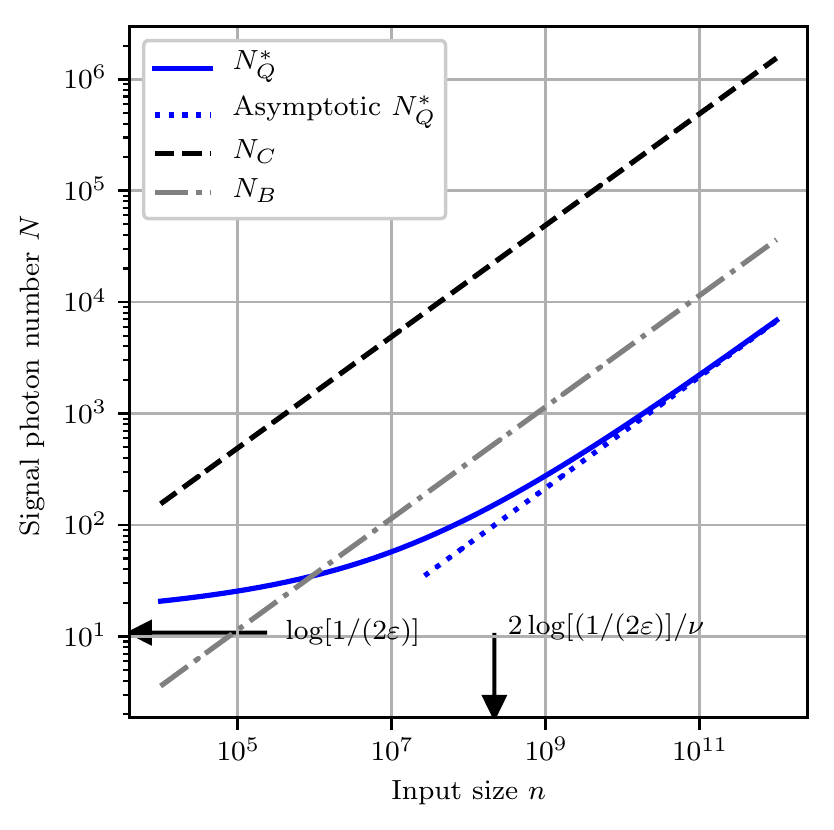}

\caption{The minimum signal photon number $N_Q^\ast$ required by the quantum fingerprinting protocol (solid line) as a function of the input size $n$ for the noise PSD $\Noise=10^{-7}$ and the desired average error probability $\varepsilon=10^{-5}$. The horizontal arrow indicates the minimum signal photon number in the noiseless scenario and the vertical arrow corresponds to the noise parameter value ${\cal N} = 1$. The dotted line is the asymptotic expression given in~(\ref{Eq:NQOptLargeNoise}). The dashed line depicts the performance of a classical fingerprinting protocol specified in~(\ref{Eq:NC}) and the dash-dotted line indicates the known classical bound given by~(\ref{Eq:NB}).}

\label{Fig:Nofn}

\end{figure}

Fig.~\ref{Fig:Nofn} compares $N_C$ and $N_B$ specified above with the numerically found minimum photon number $N_Q^\ast$ used by the quantum fingeprinting protocol for the input size $n$ in the range $10^{4} \le n \le 10^{12}$, the desired probability of error $\varepsilon=10^{-5}$, and the noise PSD $\nu = 10^{-7}$ photons per unit time and unit bandwidth. The noise parameter $\SysOp$ defined in~(\ref{Eq:NoiseParameter}) becomes equal to one for $n =  2 \Noise^{-1}\log[1/(2\varepsilon)] \approx 2.2 \times 10^{8}$. It is seen that below this threshold $N_Q^\ast$ exibits weak dependence on $n$, staying within factor of $20$ from the noiseless figure given according to~(\ref{Eq:NQNoiseless}) by
$\log[1/(2\varepsilon)] \approx 10.8$ photons. Well above the threshold corresponding to $\SysOp = 1$, the signal photon number $N_Q$ follows $O(\sqrt{n})$ scaling with the asymptotic expression (\ref{Eq:NQOptLargeNoise}) that approximates well numerical results as seen in Fig.~\ref{Fig:Nofn}.
In this regime the quantum advantage has the form of a reduced multiplicative factor compared to~(\ref{Eq:NC}) and~(\ref{Eq:NB}). The principal reason behind this reduction is distinct dependence on the AWGN strength $\nu$: the factor $1/\log_2(1+\nu^{-1})$, corresponding to the inverse of the PIE, is replaced by $\sqrt{\nu}$ in the quantum case. In the numerical example considered here with $\nu=10^{-7}$ the ratio between these two factors exceeds two orders of magnitude and it would grow further for lower $\nu$.

\section{Conclusions}
\label{Sec:Conclusions}

We have presented a theoretical analysis of a quantum fingerprinting protocol using power-limited optical signals transmitted over AWGN channels with noise strength much less than one photon per unit time and unit bandwidth. Although for large input size no scaling advantage over classical fingerprinting is retained, the quantum protocol allows one to shorten the transmission time by a multiplicative factor that depends on the noise strength. The improvement offered by quantum fingerprinting is rooted in the joint detection of the received signals. Statistics provided by such detection allows one to perform the equality test more efficiently compared to a scenario when classical fingerprints need to be recovered faithfully after signal detection. \mj{The
advantage of the quantum fingerprinting protocol over the classical one can be also phrased in terms of the amount of information about the input bit strings revealed to the referee by Alice and Bob \cite{Arrazola2014}.}

It is worth noting that joint detection used in quantum fingerprinting exploits both wave and particle properties of light: the received optical fields interfere as waves on the beam splitter, but subsequently produce discrete photocounts which at the fundamental level correspond to absorption of individual particles---photons---from light incident on photodetectors.
The process of generating photocounts by an incident electromagnetic field is inherently random. In the case of the quantum fingerprinting protocol described here, generation of a photocount by one of the photodetectors in a given temporal slot provides certain information on the phase relation between pulses transmitted in that slot. In turn, this phase relation depends on specific bits in codewords $E(x)$ and $E(y)$ encoding inputs.
Informally speaking, photon counting selects randomly, through the physics of the photodetection process, a small subset of codeword bits that are effectively compared by the referee. Signals sent by Alice and Bob are so weak that they generate photocounts only in very few slots out of their total number.

It is insightful to juxtapose the above observation with a classical fingerprinting protocol which uses shared randomness between Alice and Bob \cite{Buhrman2001}. In such a protocol Alice and Bob send only subsets of codeword bits that are specified by a shared random key. It is then sufficient to send classical fingerprints of constant length for a given probability of error. Quantum fingerprinting can be viewed as a method to replace the random key shared between Alice and Bob by the randomness of the photodetection process. In the quantum case, selection of codeword bits to be compared occurs only at the detection stage and does not require any ancillary resource to be shared between Alice and Bob.

\mj{
The quantum fingerprinting protocol described here requires setting a proper phase relation between the fields received from Alice and Bob that are interfered at the beam splitter on the referee side.
This requirement can be satisfied by transmitting additional reference signals that are measured by the referee to estimate the relative phase between the received optical fields and to adjust their phase relation with the help of a phase modulator inserted before the receiver beam splitter. Implementation of this strategy requires only a minor overhead in terms of the total transmitted optical energy, enabling one to maintain the advantage of the quantum fingerprinting protocol. To give a quantitative example, $N_{\text{est}} = 18/(\Delta \phi)^2$ photons is sufficient to estimate the relative phase with the uncertainty below $\Delta \phi$ and 99.7\% confidence  \cite{Makarov2004}. Assuming Gaussian phase fluctuations, the uncertainty $(\Delta \phi)^2$ contributes a multiplicative factor $W=\exp[-(\Delta \phi)^2/2]$ to the visibilities defined in (\ref{Eq:VeVd}). Taking for concreteness $W=0.95$ yields $N_{\text{est}} \approx 180$ photons. This figure is substantially lower than the gap between $N_Q^\ast$ and $N_B$ for the numerical example depicted in Fig.~\ref{Fig:Nofn}, in the regime $n \gg 2 \log[(1/(2\varepsilon)]/\Noise$ which corresponds to the noise parameter $\SysOp \gg 1$. Importantly, in this regime both visibilities $\Ve$ and $\Vd$ for the optimal bandwidth $\snrR^\ast$ are substantially below one, as implied by Fig.~\ref{Fig:Optimal}. Therefore, their rescaling by $W$ can be included in a straightforward manner in the approximation (\ref{Eq:Chernoffapprox1/(1+snrR)}) leading to (\ref{Eq:NQOptLargeNoise}). This produces an additional multiplicative factor $W^{-1}$ in the expression for $N_Q^\ast$ derived in (\ref{Eq:NQOptLargeNoise}). In the present example $W^{-1}\approx 1.05$ which implies that the assumed phase uncertainty does not alter noticeably $N_Q^\ast$ in Fig.~\ref{Fig:Nofn} when $\SysOp \gg 1$.}

\mj{
A practical limitation when implementing the quantum fingerprinting protocol with phase estimation described above is the number of temporal slots that can be accommodated within the coherence time of the generated optical signals. Using state-of-the-art sub-Hz linewidth lasers \cite{oewaves} and phase modulators reaching 100~GHz bandwidth \cite{Noguchi98} yields the available number of slots up to $10^{11}$. Given that the required code rate is above $0.1$ in the regime $\SysOp \gg 1$, this number of slots should be sufficient to achieve the quantum advantage for the input size $n \sim 10^{9}$--$10^{10}$ and other parameters as in Fig.~\ref{Fig:Nofn}, even when taking into account the overhead required for phase estimation. A more universal strategy, applicable also for longer inputs, is to interleave the fingerprint signal with the reference signal at intervals shorter than the coherence time so that the referee can track the relative phase between the received signals. In terms of the required optical energy, such phase tracking adds an overhead scaling linearly with the transmission time and hence proportional to $N_Q^\ast$, which retains a constant separation between $N_Q^\ast$ and $N_B$ for large input size $n$ in the logarithmic scale of Fig.~\ref{Fig:Nofn}. Yet another option to implement the quantum fingerprinting protocol is to exploit higher-order optical interference for signals without a defined phase relation \cite{Jachura2017,Jachura2018}. For this scenario, a preliminary analysis of the quantum advantage in terms of transmitted information has been recently presented \cite{Lipka2019}.
}

\mj{On an ending note, the problem of comparing weak optical signals carrying classical or quantum information occurs in a number of quantum information protocols. Two relevant classes are quantum digital signatures \cite{Clarke2012}, which provide a secure method to sign a message preventing impersonation, repudiation, or message tampering, and communication complexity protocols based on the so-called quantum switch \cite{Wei2019}.
Quantum fingerprinting can be viewed as a generic example of efficient extraction of information via optical interference and its thorough characterization may come in useful when analyzing other protocols based on a similar paradigm.}

\section*{Acknowledgments}
Insightful discussions with B. A. Bash, N. L\"{u}tkenhaus, F. Xu, and Q. Zhang are gratefully acknowledged.

\appendices

\section{}
\label{Appendix:CharacteristicFunction}

Using the orthogonality properties of the pulse mode function given in~(\ref{Eq:ModeFunction}), the integrals (\ref{Eq:Ipm}) can be brought to the form
\begin{align}
I_{\pm} &= \int_{-\infty}^{\infty} |{\cal E}^{\pm}(t)|^2 = \sum_{l=1}^{L} \left| \frac{\alpha^x_l \pm \alpha^y_l}{\sqrt{2}}\right|^2 \\
 &= \sum_{l=1}^{L} \left[ |\gamma^{\pm}_l|^2 + \left| \frac{\xi_l \pm \zeta_l}{\sqrt{2}} \right|^2 \nonumber
+ 2 \text{Re} \left( \gamma^{\pm}_l \frac{\xi_l \pm \zeta_l}{\sqrt{2}} \right)\right],
\end{align}
where
\begin{equation}
\gamma^{\pm}_l = \sqrt{\frac{\Signal}{2B}} (e^{i\theta_l^x}\pm e^{i\theta_l^y}).
\end{equation}
Note that linear combinations $(\xi_l \pm \zeta_l)/\sqrt{2}$ are Gaussian random variables with zero mean and variance $\text{Var}[ (\xi_l \pm \zeta_l)/\sqrt{2}] = \Noise$. This allows one to calculate directly the expectation value in~(\ref{Eq:Zlambda+lambda-}) which yields:
\begin{align}
Z(\lambda_+, \lambda_-)
&=  \exp \Bigg[ (e^{i\lambda_+}-1) 
\left( 1 + \frac{(e^{i\lambda_+}-1)\nu}{1-(e^{i\lambda_+}-1)\nu} \right)
\sum_{l=1}^{L} |\gamma_l^+|^2 \nonumber \\
&+ ( e^{i\lambda_-}-1) \left( 1 + \frac{(e^{i\lambda_-}-1)\nu}{1-(e^{i\lambda_-}-1)\nu} \right)
\sum_{l=1}^{L} |\gamma_l^-|^2 \Bigg] \nonumber
\\
&\times \left(\frac{1}{1-(e^{i\lambda_+}-1)\nu}\right)^L
\left(\frac{1}{1-(e^{i\lambda_-}-1)\nu}\right)^L.
\label{Eq:Zafterintegr}
\end{align}
The terms in the exponent involving $\nu$ produce expressions of the order $O(\nu LS/B)$ and will be neglected. Sums over $l$ can be written as
\begin{equation}
\sum_{l=1}^{L} |\gamma_l^\pm|^2 = \frac{LS}{B} \left( 1\pm \frac{1}{L} \sum_{l=1}^{L} \cos (\theta_l^x- \theta_l^y) \right).
\end{equation}
Furthermore, for $\nu \ll 1$ and large $L$ the power factors in~(\ref{Eq:Zafterintegr}) can be approximated by exponents $1/[1-(e^{i\lambda_\pm}-1)\nu]^L \approx \exp[(e^{i\lambda_\pm}-1)\nu L]$. Combining these steps together yields
\begin{multline}
Z(\lambda_+, \lambda_-) =
\exp\Bigg[ (e^{i\lambda_+}-1) L 
\left(\frac{S}{B} + \nu + \frac{1}{L} \sum_{l=1}^{L} \cos (\theta_l^x- \theta_l^y) \right) \\
+ (e^{i\lambda_-}-1) L \left(\frac{S}{B} + \nu - \frac{1}{L} \sum_{l=1}^{L} \cos (\theta_l^x- \theta_l^y) \right) \Bigg]
\end{multline}
which is identical with~(\ref{Eq:Zfinal}) when expressed in terms of $\mu$ and $V$ defined respectively in~(\ref{Eq:mudef}) and (\ref{Eq:vgen}).

\section{}
\label{Sec:AppB}

The argument $\lambda^\ast$ optimizing the right hand side of~(\ref{Eq:ChernoffPerClick}) can be found by solving equation $df/d\lambda=0$, where
\begin{equation}
f(\lambda)=1-\frac{1}{2}[(1+\Vise)^{\lambda}(1+\Visd)^{1-\lambda}
+(1-\Vise)^{\lambda}(1-\Visd)^{1-\lambda}].
\label{Eq:flambda}
\end{equation}
The solution is given by the following closed expression:
\begin{equation}
\lambda^\ast=\frac{\log\left[\frac{1-\Visd}{1+\Visd}\frac{\log\frac{1-\Visd}{1-\Vise}}{\log\frac{1+\Vise}{1+\Visd}}\right]}{\log\left(\frac{1+\Vise}{1-\Vise}\frac{1-\Visd}{1+\Visd}\right)}.
\end{equation}
For  $\Vise,\Visd \ll 1$ the above formula can be approximated up to the second order by
\begin{equation}
\lambda^\ast\approx\frac{1}{2}+\frac{\Visd^2-\Vise^2}{24}.
\end{equation}
Inserting this expression into~(\ref{Eq:flambda}) yields up to the second order in $\Vise,\Visd$:
\begin{equation}
\textsf{C}\approx\frac{1}{8}\left(\Vise-\Visd\right)^2 .
\end{equation}
The same result is obtained by using the zeroth order expansion $\lambda^\ast \approx 1/2$ in~(\ref{Eq:flambda}).

\bibliographystyle{IEEEtran}
\bibliography{lipka}
%\printbibliography


\begin{IEEEbiography}[{\includegraphics[width=1in,height=1.25in,clip,keepaspectratio]{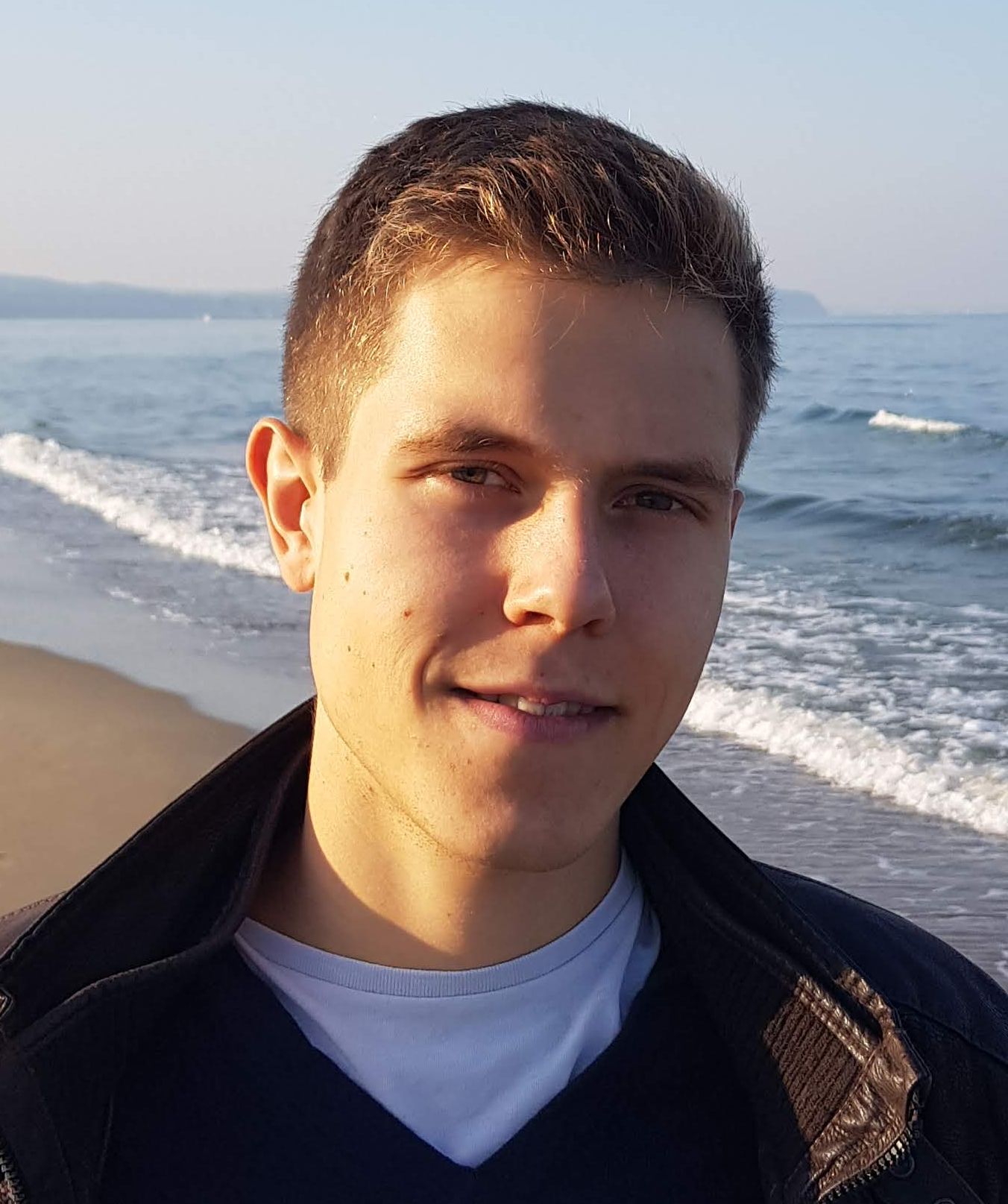}}]{Michał Lipka}
Michał Lipka received the B.Sc. degree in physics from the University of Warsaw, Poland in 2018.

He is doing his M.Sc in physics at the University of Warsaw and since 2015 works there in the Quantum Memories Laboratory which is now a part of the Centre for Quantum Optical Technologies.

In 2019 he has been awarded National (Poland) Ministry of Science and Higher Education's "Diamond Grant" to develop real-time single photon localization technologies. 
\end{IEEEbiography}

\begin{IEEEbiography}[{\includegraphics[width=1in,height=1.25in,clip,keepaspectratio]{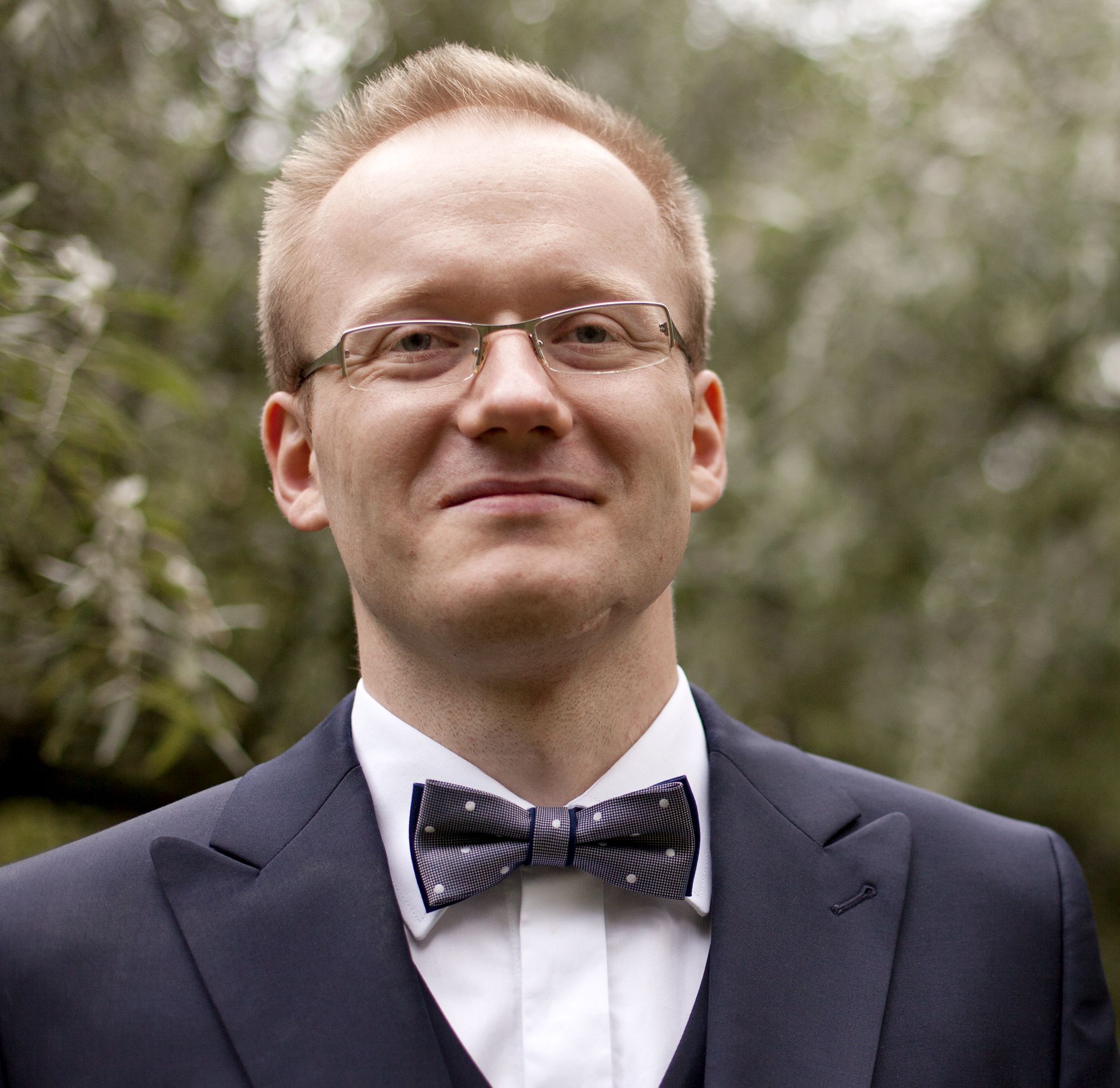}}]{Marcin Jarzyna}
Marcin Jarzyna received the M.Sc. and Ph. D. degrees in physics from the University of Warsaw, Warsaw, Poland, in 2011 and 2016 respectively. 

His research was focused mainly on quantum metrology and the impact of entanglement on the asymptotic precision limits under decoherence. He has been working at the Centre of New Technologies, University of Warsaw since 2016.

 His current research interests include low power limits of communication, impact of signal amplification, superresolution effects in optical imaging and optical realizations of communication complexity problems
\end{IEEEbiography}

\begin{IEEEbiography}[{\includegraphics[width=1in,height=1.25in,clip,keepaspectratio]{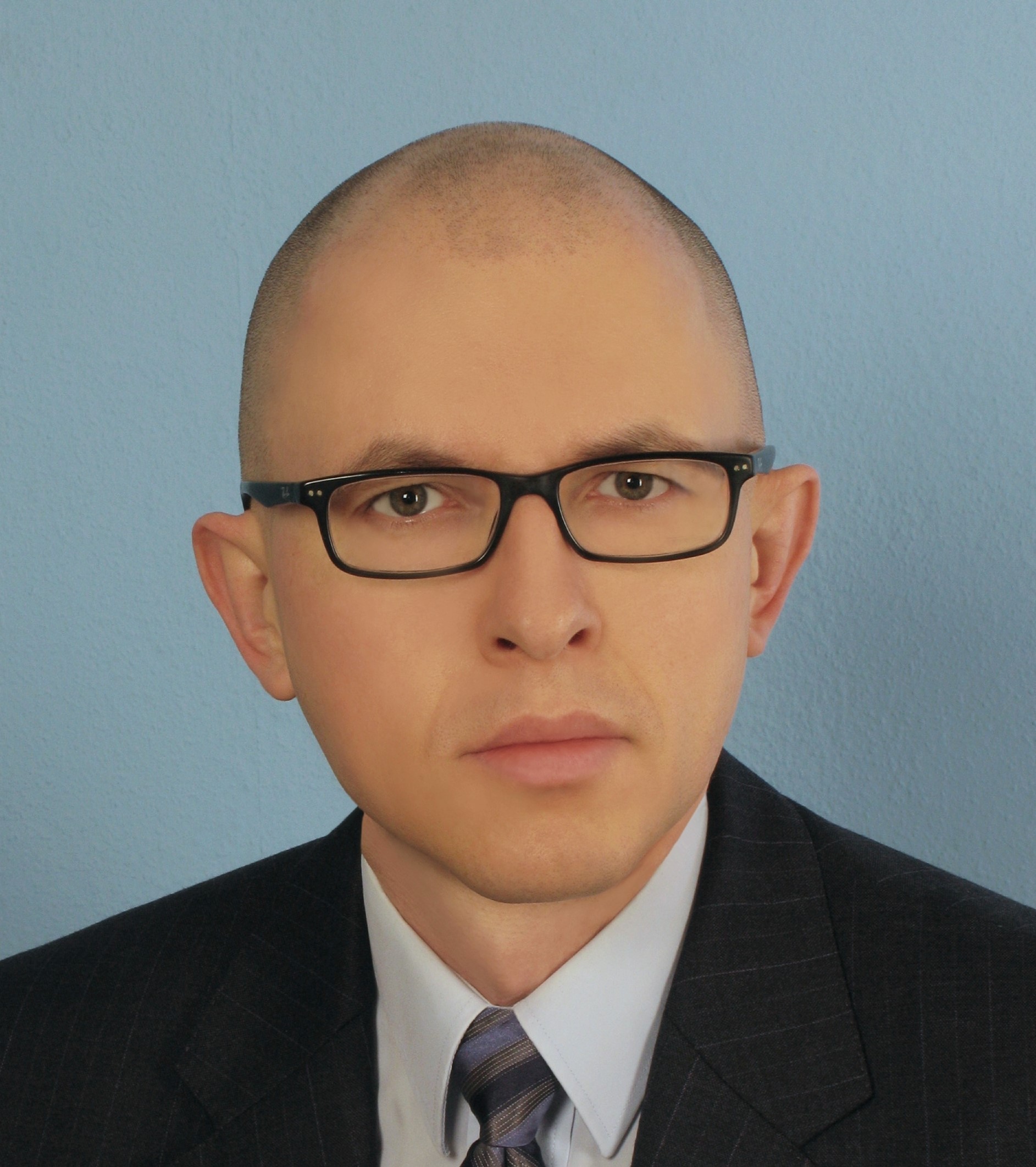}}]{Konrad Banaszek}
Konrad Banaszek received the M.Sc. and Ph.D. degrees in physics from the University of Warsaw, Poland, respectively in 1997 and 2000. 

He held postdoctoral positions at the University of Rochester, NY and the University of Oxford, UK, followed by a Junior Research Fellowship at St. John's College, Oxford, UK and faculty appointments at the Nicolaus Copernicus University in Toru\'{n}, Poland, from 2005 to 2009, and at the University of Warsaw, Poland, since 2009. 
His field of research is quantum physics and optical sciences with a focus on novel approaches to communication, sensing, and imaging that enable operation beyond the standard quantum limits. Currently he is the director of the Centre for Quantum Optical Technologies established in 2018 by the University of Warsaw in partnership with the University of Oxford under the International Research Agendas Programme operated by the Foundation for Polish Science.

Dr. Banaszek has served as an Associate Editor of Optics Express and a guest editor of a focus issue of New Journal of Physics on quantum tomography. In 2001 he received the European Physical Society Fresnel Prize for his contributions to the understanding of non-classical light and its applications in quantum information processing.

\enlargethispage{-0.5in}
\end{IEEEbiography}


%\newpage



%\vfill


\end{document}